\documentclass[manuscript,nonacm]{acmart}

\usepackage{enumitem}
\usepackage{tikz}
\usepackage{array}
\usepackage{wrapfig}

\usepackage{listings}
\usepackage{xcolor}

\lstdefinestyle{prompt}{
  basicstyle=\ttfamily\footnotesize,
  breaklines=true,
  breakindent=0pt,
  columns=fullflexible,
  keepspaces=true,
  frame=single,
  rulecolor=\color{black!30},
  backgroundcolor=\color{black!4},
  xleftmargin=4pt,
  xrightmargin=4pt,
  aboveskip=6pt,
  belowskip=6pt,
  showstringspaces=false,
}
\lstnewenvironment{promptlisting}{\lstset{style=prompt}}{}

\usetikzlibrary{external, shapes.geometric, arrows}
\usepackage{xspace}
\acmDOI{}          
\acmISBN{}
 \title[short]{full}

\AtBeginDocument{%
  \providecommand\BibTeX{{%
    \normalfont B\kern-0.5em{\scshape i\kern-0.25em b}\kern-0.8em\TeX}}}

\begin{document}

\tikzstyle{5_box_node} = [
    rectangle,
    rounded corners, 
    minimum width=2cm, 
    minimum height=1cm,
    text centered,
    text width = 2.5cm,
    draw=black,
]
\tikzstyle{3_box_node} = [
    rectangle,
    rounded corners, 
    minimum width=3cm, 
    minimum height=1cm,
    text centered,
    text width = 4cm,
    draw=black,
]
\tikzstyle{4_box_node} = [
    rectangle,
    rounded corners, 
    minimum width=3cm, 
    minimum height=1cm,
    text centered,
    text width = 3.2cm,
    draw=black,
]
\tikzstyle{arrow} = [thick,->,>=stealth]


\newcommand{\yaqing}[1]{\textcolor{violet}{#1}}
\newcommand{\sssec}[1]{\vspace*{0.05in}\noindent\textbf{#1}}
\newcommand{\dnote}[1]{\textcolor{blue}{$\langle$#1 -- Dafna$\rangle$}}

\renewcommand\footnotetextcopyrightpermission[1]{}

\newcommand{\sys}{\text{StructSim}\xspace}

\title{\sys: Measuring Idea Similarity at Scale Through Structural Representation}

\author{Yaqing Yang}
\affiliation{%
  \institution{Carnegie Mellon University}
  \city{Pittsburgh, PA}
  \country{USA}}
\email{yaqingyy@andrew.cmu.edu}

\author{Vikram Mohanty}
\affiliation{%
  \institution{Carnegie Mellon University}
  \city{Pittsburgh, PA}
  \country{USA}}
\email{vikrammohanty@acm.org}

\author{Mei-Xi Chia}
\affiliation{%
  \institution{Carnegie Mellon University}
  \city{Pittsburgh, PA}
  \country{USA}}
\email{mchia@andrew.cmu.edu}

\author{Nikolas Martelaro}
\affiliation{%
  \institution{Carnegie Mellon University}
  \city{Pittsburgh, PA}
  \country{USA}}
\email{nikmart@cmu.edu}

\author{Dafna Shahaf}
\affiliation{%
  \institution{The Hebrew University of Jerusalem}
  \city{Jerusalem, NO}
  \country{Israel}}
\email{dshahaf@cs.huji.ac.il}

\author{Aniket Kittur}
\affiliation{%
  \institution{Carnegie Mellon University}
  \city{Pittsburgh, PA}
  \country{USA}}
\email{nkittur@cs.cmu.edu}

\renewcommand{\shortauthors}{Yang et al.}

\begin{abstract}






Measuring \emph{idea similarity} is fundamental to creativity evaluation, especially as LLMs enable idea generation at increasing scale. However, text embeddings collapse an idea into a single vector, making it difficult to capture structural similarity, including partial overlap across core and supporting components and differences across levels of abstraction. We introduce a shared structural representation that decomposes ideas into purpose, mechanism, and implementation components and organizes related components in a multi-layer concept graph. From this representation, we define measures of pairwise similarity and set-level mechanism coverage for assessing idea diversity. We evaluate our approach using controlled idea triples and assessments from 12 experts, focusing on differences in core mechanisms, implementations, and supporting components. Our method improves alignment with expert judgments of structural similarity by 31\% over the embedding baseline and better reflects expert assessments of idea set coverage, supporting scalable evaluation of idea similarity and diversity.
\end{abstract}


\maketitle

\section{Introduction}
\label{intro}

\begin{figure}[h]
\centering
\includegraphics[width=\linewidth]{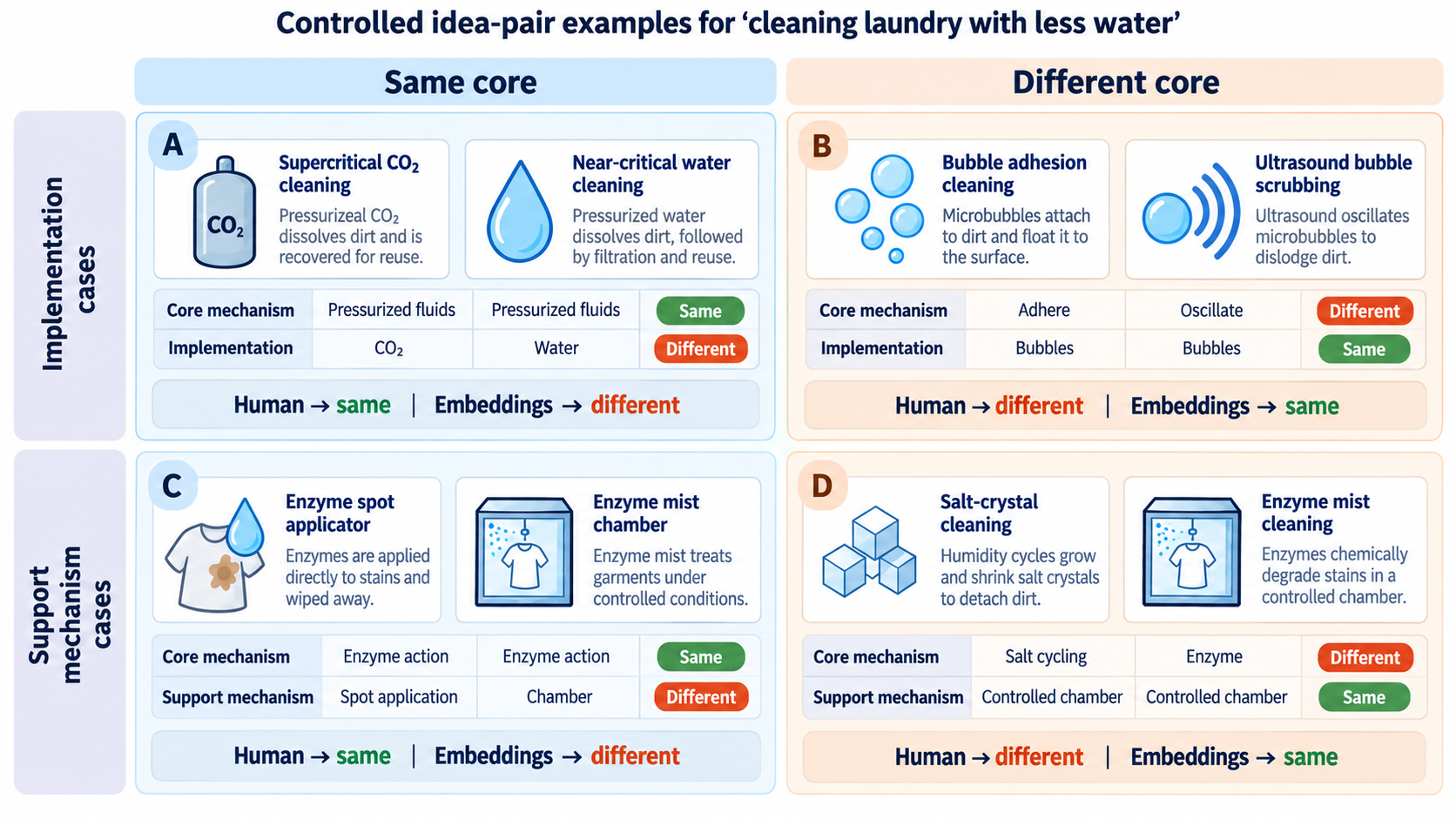}
\caption{Examples of idea similarity based on idea structure. We distinguish core from supporting parts and higher-level purpose and mechanism from lower-level implementation. All ideas address the problem of cleaning laundry with less water. In the left column (A, C), each pair shares the same core mechanism but differs in implementation. People therefore judge these ideas as similar, while embeddings may overemphasize differences in wording or implementation. In the right column (B, D), each pair uses different core mechanisms but shares an implementation or supporting feature. People judge these ideas as different, while pure embedding-based method may score them as similar because of these surface similarities.}
\label{fig:idea_sim}
\Description{The figure shows four controlled idea pairs for cleaning laundry with less water. It contrasts cases where ideas share the same core mechanism but differ in implementation or support with cases where they share surface features but use different core mechanisms, illustrating where human and embedding-based similarity judgments may diverge.
}
\end{figure}

Large language models (LLMs) have fundamentally shifted the landscape of creative idea generation; however, our ability to understand the ways they will be beneficial or harmful will require a robust way to measure the ideas generated by those models. Some studies measure LLMs as generating ideas that are as good as or better than
humans'~\cite{si2025can,koivisto2023best,hubert2024current} and, with the right prompting, as diverse~\cite{meincke2024prompting,feng2025enhancing,hu2024novaiterativeplanningsearch}. Others measure the opposite: LLM ideas as being
homogeneous~\cite{ma2024diverse,wenger2026homogeneously,anderson2024homogenization,doshi2024generative,zhang2025noveltybench} with surface novelty disappearing on closer scrutiny~\cite{chakrabarty2024art}. 

Furthermore, the importance of similarity measurement goes beyond being an offline evaluation metric; it is the foundational computational infrastructure required to unlock true creativity support. Creativity support tools whose core operations depend on measures of automated similarity judgments, such as clustering, deduplication, and ``show me something different,'' are meant to expand a creator's range and
agency~\cite{shneiderman2007creativity,deterding2017mixed}; however when that measure fails to capture the deep structure of the idea space the same operations can instead fixate, feeding users variations on one idea~\cite{,anderson2024homogenization,doshi2024generative}. Lack of well grounded similarity metrics can also have downstream effects on AI models that assume them as truth. Generative models and ideation algorithms are being evaluated, benchmarked, and RLHF-optimized against metrics that actively penalize deep structural novelty in favor of lexical variance~\cite{hu2024novaiterativeplanningsearch,lanchantin2025diversepreferenceoptimization,kirk2024understandingeffectsrlhfllm,shaib2025standardizing}.


Yet, despite its centrality to how we evaluate and build systems, the field currently lacks an accurate, scalable, and reliable instrument capable of distinguishing ideas that share an underlying approach from those that contribute meaningfully different ones~\cite{shah2003metrics}. 

One recent class of approaches that has gained popularity are LLM-as-a-Judge approaches, prompting models to directly assess the functional or conceptual relationships between ideas~\cite{lu2024aiscientistfullyautomated,li2025chainideas,baek2025researchagent,radensky2026scideator,zhang2025exploring}. However, such approaches have several limitations that make them less appealing for widespread use as a similarity metric. Their evaluations can be stochastic and inconsistently aligned with human judgments~\cite{zhang2025exploring,li2025chainideas, chakrabarty2024art}, and scale poorly in terms of accuracy and cost with hundred or thousands of ideas~\cite{meincke2024prompting}. Perhaps most concerning, they are inherently unstable metrics, potentially giving different results on each run (even when holding factors such as temperature and seed constant) and when items are added or reordered, ~\cite{atil2024stability,stureborg2024inconsistent,wang2023notfair,shi2024positionbias,zheng2023judging}. 

In part due to the above issues, dense vector embeddings (e.g., Sentence-BERT, OpenAI text-embeddings) scored via cosine distance remain the dominant instrument for assessing whether LLMs generate diverse concepts~\cite{si2025can,hu2024novaiterativeplanningsearch,anderson2024homogenization,meincke2024prompting,doshi2024generative,ashkinaze2025howai,wenger2026homogeneously,kirk2024understandingeffectsrlhfllm,ma2023conceptual,ma2024diverse,feng2025enhancing}. However, because the semantic meaning captured by embeddings does not necessarily reflect the deep structural core mechanisms of an idea~\cite{fodor2025compositionality,nikolaev2023representation}, 
they could fail to bridge the classic gap between surface syntax and structural mechanisms~\cite{gentner1983structure,gentner1993roles}. Specifically, such measure can fail to distinguish between surface-level syntax and underlying structural mechanisms, and between mechanisms that are core to the central idea versus more peripheral ideas that support that core.  Figure~\ref{fig:idea_sim} shows pairs of ideas for the problem of cleaning laundry with less water that illustrate how these distinctions can lead to different judgments of similarity between humans and machines. In pair (a), pressurized CO$_2$ and near-critical water share the same core mechanism of cleaning with pressurized fluids but differ in implementation; as we show later, computational approaches that rely on surface vocabulary focus on the latter while human experts align on the former. Likewise, pair (c) shares the same core mechanism of enzyme action but differ in its supporting mechanism; humans focus on the core while common scalable approaches such as embeddings are fixate on the latter. The converse can result in human-machine differences in judgment as well: when surface-level implementions or supporting mechanisms are similar but core mechanisms differ.

In summary, the field is currently lacking a viable instrument for measuring similarity: human judgments cannot scale over combinatorial idea spaces; dense vector embeddings scale seamlessly but remain blind to structural mechanism; and LLM judges discern mechanism but remain unstable, opaque, uncacheable, and computationally prohibitive.

To address these issues, we introduce \textbf{\sys}, a framework that combines sensitivity to structural mechanisms with scalability, and incremental reusability. Rather than purely relying on LLM judgments at runtime, \sys uses a single structured LLM pass to analyze each idea, requiring $O(n)$ total LLM calls. The LLM first decomposes each idea into multiple parts, distinguishing core from supporting parts. Each part is also represented as a three-layer graph that captures its structural logic at different levels of abstraction, from high-level purpose and mechanism to low-level implementation. Concepts at each layer can then be independently encoded in the embedding space. Across the idea corpus, \sys clusters similar concepts within each layer to construct a global concept graph. Once this graph is built, subsequent similarity evaluations require no additional LLM calls and are computed using lightweight, deterministic vector operations over the concept graph.

By shifting the LLM's role from a costly runtime judge to a single-pass structural feature generator, 
\sys provides:

\begin{itemize}
  \item \textbf{Mechanism-Aware Accuracy:} Differentiates surface prose from underlying functional logic, better matching expert human judgments of structural similarity.
  \item \textbf{Deterministic and Incremental Scaling:} Generates stable, cacheable point representations ($O(1)$ comparisons) that allow new ideas to be indexed against an existing canvas without re-evaluating prior concepts.
  \item \textbf{Inspectable and Steerable Interactions:} Provides an explicit structural geometry, enabling creators to filter and navigate design spaces at varying abstraction levels (e.g., ``find concepts with the same functional goal but a distinct execution mechanism'').
\end{itemize}

Through rigorous empirical evaluations, we demonstrate that our approach addresses the measurement misalignment of standard embeddings and exceeds embedding baselines in alignment with expert judgments. By establishing a scalable, mechanism-aware measure of similarity, our work takes a step toward providing a more stable foundation for creative AI evaluation and the structural infrastructure needed to build more useful co-creative tools. This paper makes the following contributions:

\begin{itemize}
  \item We demonstrate that standard embeddings systematically mismeasure idea diversity, showing how embedding-selected diverse sets systematically differ from expert judgments
  \item We introduce a single-pass decomposition framework that constructs hierarchical trees of vector embeddings to represent concepts across multi-level structural abstractions.
\item We empirically characterize where embedding-based idea similarity diverges from expert judgments, focusing on differences in core mechanisms, implementations, and supporting components. Our method improved alignment with human judgments of structural similarity by 31\% over the our best embedding baseline.
\end{itemize}
\section{Related Work}
\label{related}

\subsection{Structural Similarity in Design Ideas}
\label{related:human}

Meaningful idea similarity depends on structural rather than literal overlap.
Prior studies of similarity show that people compare concepts based on shared and different properties and relations, rather than only on surface resemblance~\cite{tversky1977features,markman1993splitting,markman1996commonalities}. This distinction is particularly important in design, where experts attend more to functional and relational structure than novices, who rely more heavily on surface features~\cite{casakin1999expertise,casakin2004visual,ozkan2013cognitive}. This focus on structural similarity matters for creativity: differences between the underlying approaches of design concepts can affect the novelty and creativity of ideas generated from them~\cite{nagai2009concept,jang2014effect}. 

To construct a reliable similarity measure, prior theories suggest that structural similarity between ideas should account for composition, structural role, and level of abstraction.
First, ideas are \emph{compositional}: similarity can arise from partial overlap on multiple components or a local structure~\cite{tversky1977features,markman1996commonalities}. Capturing \emph{which components} overlap is important for idea evaluation, as the same overall similarity score can describe very different relationships. Distinguishing these cases helps determine whether a new idea extends an existing approach or contributes a meaningfully different direction, which is central to assessing novelty and diversity~\cite{shah2003metrics}.
Second, idea components can differ in their \emph{structural role}; theories of feature salience and systematicity suggest that structurally important commonalities should contribute more to similarity than peripheral ones~\cite{tversky1977features,clement1991systematicity}. Third, design ideas can be represented at different \emph{levels of abstraction}. Across design frameworks, a recurring distinction separates what a design aims to achieve, the principle through which it works, and its concrete realization. For example, FBS distinguishes function, behavior, and structure~\cite{qian1996function}; McTeague et al. distinguish need, function, and means~\cite{mcteague2018insights}; and Shah et al. distinguish physical and working principles from embodiment and detail when measuring idea variety~\cite{shah2003metrics}. Shah et al. also treat differences at higher structural levels as more substantial than differences in lower-level embodiment or detail, supporting the need to distinguish the level at which ideas overlap~\cite{shah2003metrics}. 

However, these theoretical frameworks do not provide a systematic way to integrate partial overlap, structural importance, and levels of abstraction into a single similarity measure, nor do they offer a computational approach for measuring structural similarity at scale. Our goal is therefore to develop a scalable structural similarity measure that operationalizes these properties within a shared computational representation of the idea space.

\subsection{Computational Similarity Assessment}
Scalable text-based similarity measures have progressed from surface-form overlap toward richer semantic comparison, but they still provide limited support for structural similarity. Early approaches measure lexical or syntactic overlap~\cite{papineni2002bleu,lin2004rouge}. These measures capture differences in wording and linguistic form~\cite{guo2025benchmarking,shypula2025evaluating}, but such differences do not necessarily reflect differences in how ideas work.
Embedding-based approaches address this limitation by comparing semantic rather than only lexical similarity. They represent ideas in a semantic vector space and compute similarity from vector proximity~\cite{reimers2019sentence}, and are widely used for similarity and diversity assessment~\cite{beaty2022semantic,ma2023conceptual,jiang2026artificial,friedman2022vendi}. However, prior work shows that embedding similarity does not always align with human-perceived similarity~\cite{tevet2021evaluating,zhou2022problems,opitz2025interpretable}. More importantly, it is unclear what information in the idea drives the resulting similarity: a single embedding does not reveal which components overlap, whether those components are core or supporting, or whether the overlap occurs at the mechanism or implementation level.

Augmented embedding approaches partly address this opacity by intentionally emphasizing information relevant to a comparison. Instruction-conditioned embeddings incorporate additional context into the representation~\cite{su2023one}, while other methods extract relevant keyphrases and combine their embeddings with the original text embedding~\cite{viswanathan2024large}. These methods provide more control over what information is emphasized, but they still do not explicitly represent how different parts of two ideas partially overlap or support flexible comparison across those parts. Moreover, selecting what information to emphasize does not by itself identify which components reflect the deeper structure of an idea rather than its surface features.

Other methods move beyond fixed embedding similarity by learning or reasoning about the comparison directly. Supervised approaches can learn a particular notion of similarity; for example, NoveltyBench trains a classifier from human annotations to identify functional equivalence~\cite{zhang2025noveltybench}. Such methods can target a specific criterion, but they require labeled data and may not generalize when the relevant notion of similarity changes across tasks or domains. LLM-based evaluation offers greater flexibility by reasoning directly about idea content~\cite{shypula2025evaluating,zheng2023judging}. However, it typically produces a separate judgment for each comparison rather than a shared representation of how ideas relate, making it difficult to support scalable and systematic comparison across an idea space.

Overall, existing computational methods progressively improve what information similarity measures can capture or emphasize, but they do not provide a shared structural representation that supports partial overlap, structural role, and level of abstraction. These properties are needed to compare the deeper structural similarity of ideas while flexibly distinguishing similarities across different parts of their structure.

\subsection{Computational Structural Representations of Ideas}

Prior work has also developed computational structural representations that make the internal organization of ideas explicit in a form that can support automated comparison, retrieval, and evaluation. One line of work relies on predefined structures. Function-based representations use standardized vocabularies and roles to describe what a design does and how it is realized~\cite{van2005model}; Design Structure Matrices represent predefined system components and their dependencies~\cite{eppinger2012design}; and design-by-analogy metrics compare products through manually represented functions and customer needs~\cite{mcadams2002quantitative}. However, constructing and applying these predefined categories often requires substantial manual effort, and the resulting structures may be difficult to generalize across ideas from different tasks and contexts. More recent approaches induce structure directly from text collections, making the representation more flexible and scalable. TechNet constructs a semantic network of technical concepts from patent data~\cite{sarica2020technet}, and TnT-LLM automatically builds interpretable taxonomies over text collections~\cite{wan2024tnt}. Graph-based approaches expose finer-grained structure: GraphEval decomposes ideas into related viewpoint nodes~\cite{feng2025grapheval}, and Functional Concept Graphs extract purposes and mechanisms from patents and organize them through abstraction relations~\cite{sweed2025finding}. These approaches can reveal meaningful structure within a corpus, but they do not systematically make explicit which idea components partially overlap, the structural role of those components, or the level of abstraction at which the overlap occurs. This limits their ability to support the structural similarity assessment we seek.

Overall, existing computational structural representations do not systematically capture all three properties needed for structural similarity: partial overlap, structural role, and level of abstraction, limiting their use for large-scale structual idea similarity assessment.
\section{Method}

\begin{figure}[h]
\centering
\includegraphics[width=\linewidth]{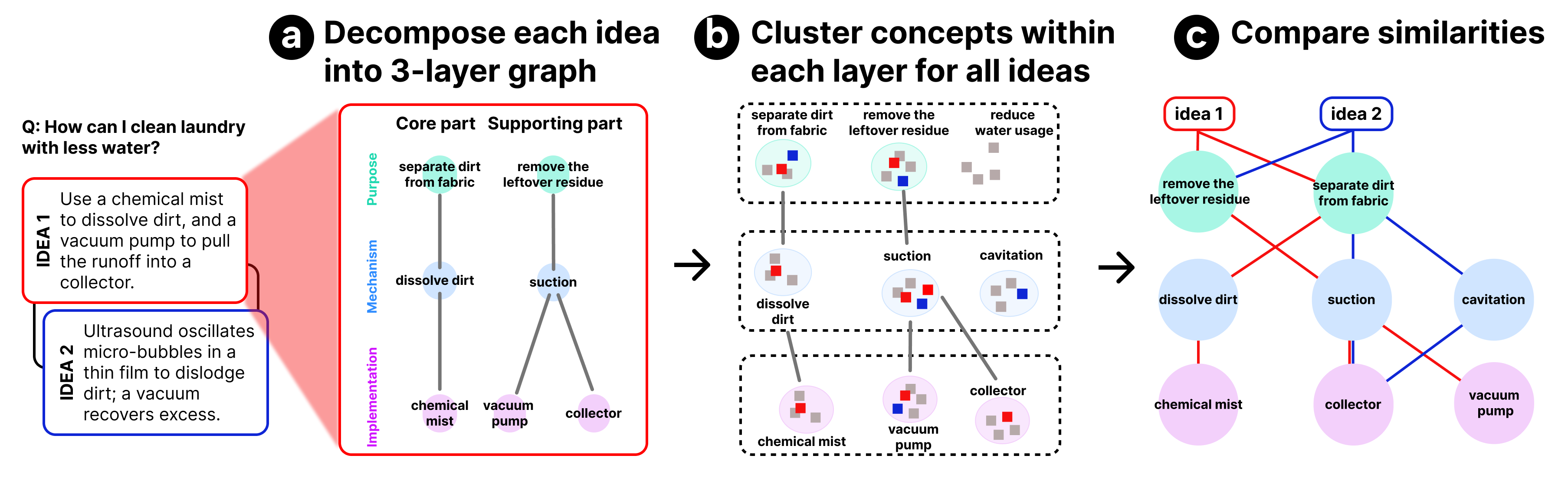}
\caption{Overview of StructSim. (a) We decompose each idea into core and supporting parts and represent each part at the purpose, mechanism, and implementation levels, so similarity can reflect partial overlap, structural importance, and differences in abstraction rather than collapsing the idea into one representation. (b) Components from all ideas are clustered separately within each layer to form a concept graph, allowing differently worded components to be compared through a common concept space while preserving how concepts connect across layers. (c) We compare ideas by matching their components in this shared space, weighting core parts and mechanism more strongly so that similarity is driven more by the idea’s structural logic than by peripheral or surface-level overlap.}
\label{fig:method}
\Description{}
\end{figure}

Our goal is to develop a scalable method for measuring structural idea similarity beyond surface wording and features. Figure~\ref{fig:method} demonstrates our pipeline for constructing the idea-space representation used for structural similarity comparison. The representation should capture partial overlap across idea parts, distinguish core from supporting parts through different weights, and represent ideas at multiple levels of abstraction, from purpose and mechanism to implementation. This structure supports both pairwise similarity assessment and set-level diversity measurement.

\subsection{Building the Idea Space}

\subsubsection{Idea decomposition to construct idea graphs for individual ideas.}
To capture the three properties important for structural similarity, we first decompose each idea into multiple parts. This decomposition allows different parts of an idea to be represented separately, assigns different importance to parts based on their structural roles, and represents each part at different levels of abstraction as shown in Figure~\ref{fig:method} a.
Inspired by prior work, each idea is decomposed into parts where each part is a three-layer graph structure to represent idea component from higher-level goal and mechanism, and lower-level implementation~\cite{mcteague2018insights,shah2003metrics,sweed2025finding}: (1) \textbf{purposes}, the goals or sub-problems addressed by the part; (2) \textbf{mechanisms}, how those purposes are achieved; and (3) \textbf{implementations}, the concrete devices, materials, techniques, or procedures used to realize the mechanisms. See prompt used in Appendix~\ref{app:prompt-decomposition}.

First, representing an idea as multiple parts allows different functional parts to be captured separately rather than collapsed into a single description. For example, consider a laundry-cleaning idea in which \emph{ultrasound vibrates microbubbles to dislodge dirt from fabric, followed by an air-blowing step that dries the fabric}. The idea contains two parts that serve different functions: the cleaning part uses ultrasonic vibration to \emph{dislodge dirt}, while the drying part uses air blowing to \emph{dry the fabric}. Separating these parts also allows us to distinguish their structural roles. In this example, the cleaning part is the \textbf{core part} because its mechanism, ultrasonic vibration, directly performs the main cleaning function, whereas the drying part is a \textbf{supporting part} that supports the overall process after the dirt has been released. Treating these parts as equally important would therefore miss an important distinction in the idea's structure.

The layered representation further distinguishes structural similarity across different levels of abstraction. For example, two ideas may both use \emph{microbubbles} as an implementation while relying on different mechanisms: Bubble adhesion cleaning'' uses bubbles to attach to and float dirt away, whereas Ultrasound bubble scrubbing'' oscillates bubbles to dislodge dirt. Conversely, two ideas can share a mechanism while differing in implementation. For instance, two water-saving ideas may both use the mechanism of \emph{recovering the cleaning medium for reuse}, while one implements this through filtration and the other through a different recovery technique. Representing purposes, mechanisms, and implementations separately therefore allows the method to identify where two ideas overlap structurally, rather than treating shared wording or a shared implementation as evidence of overall similarity.



\subsubsection{Concept clustering to construct a global concept graph for all ideas.}
To make the decomposed ideas comparable, we organize semantically similar decomposed items into shared concepts as shown in Figure~\ref{fig:method} b. Items are embedded and clustered separately within each layer using agglomerative clustering with cosine distance. Rather than using a fixed distance threshold, we set the clustering threshold $\theta_l$ to the $q$-quantile of the observed pairwise distance distribution within layer $l$. We used q=0.05 in our experiment. This allows the concept vocabulary to adapt to design problems with different levels of semantic variation without requiring a predefined taxonomy. Each resulting cluster defines a concept node represented by the normalized mean embedding of its members.

\paragraph{Soft concept membership.}
A decomposed item may fall between several related concepts, so we avoid assigning it to only one cluster. Instead, each item is connected to its $K$
nearest concepts with softmax-normalized weights. We set K to 3, and \(\tau=0.05\) to produce relatively sharp soft assignments while retaining nonzero membership for nearby concepts; because cosine similarities are normalized to a common bounded scale, the same temperature can be applied consistently across tasks without task-specific tuning. Within each layer, for item x and any concept c with centroid $\mu_c$, we have:

\begin{equation}
\label{equ: softmax}
w(x,c) =
\frac{\exp\!\big(\cos(x,\mu_c)/\tau\big)}
{\sum_{c' \in \mathrm{top}\text{-}K(x)}
\exp\!\big(\cos(x,\mu_{c'})/\tau\big)}.
\end{equation}

This makes the representation less sensitive to small differences near
clustering boundaries. Aggregating these memberships across the items of idea
$i$ gives its concept distribution in layer $l$, $p_l(i)[c]$.

\paragraph{Cross-layer structure.}
To maintain the relations of concepts across different layers, we also map the links identified in the individual idea graph into the global concept graph space.
Purpose--mechanism and mechanism--implementation links are weighted by the
memberships of their endpoints. We retain these links because two ideas may
contain similar concepts but connect them in different ways.

\paragraph{Concept rarity.}
Concepts that occur in most ideas in one idea set provide little information for distinguishing ideas. We therefore downweight common concepts using inverse document frequency. 
\begin{equation}
\mathrm{idf}(c) = \frac{1}{\log_2(1+\mathrm{df}(c))},
\end{equation}
where $\mathrm{df}(c)$ is the number of ideas with meaningful membership in
concept $c$. Cross-layer edges are weighted in the same way.

\subsection{Pairwise similarity.}
Using the soft concept memberships defined above, we calculate idea similarity by matching components in this shared space, weighting core parts and mechanisms more strongly so that similarity is driven by the idea’s structural logic rather than peripheral or surface-level overlap.
We first represent each decomposed item as a sparse vector over the concepts in its layer. For example, suppose a mechanism item is most closely associated with three concepts, $c_1$, $c_2$, and $c_3$, with soft membership weights $0.75$, $0.20$, and $0.05$. When we set $K=3$ in~\ref{equ: softmax}, its concept-membership vector is therefore $[0.75, 0.20, 0.05]$, where each dimension corresponds to a concept in the shared vocabulary rather than to a dimension of the original embedding. This representation allows two items with different wording to be similar when they are associated with the same underlying concepts.

We compute pairwise idea similarity separately for the purpose, mechanism, and implementation layers. Within each layer, we use symmetric mean-max matching over these item vectors. For each item in idea $i$, we find the item in idea $j$ with the highest cosine similarity and average these best-match scores; we then repeat the process from $j$ to $i$ and average the two directions. To give greater importance to the central content of each idea, the decomposition marks one purpose and one mechanism as core. These core items receive weight $\gamma$, while all remaining items receive weight $1$, with the weights normalized within each idea. Specifically, for layer $l$,

\begin{equation}
S_l(i,j) =
\frac{1}{2}
\left[
\frac{\sum_x w_x \max_y \cos(x,y)}{\sum_x w_x}
+
\frac{\sum_y w_y \max_x \cos(y,x)}{\sum_y w_y}
\right],
\end{equation}

where $w_x=\gamma$ for core items and $w_x=1$ otherwise. Cross-layer similarity is computed using IDF-weighted Jaccard similarity over the concept-level edges defined above.

The final similarity is:

\begin{equation}
S(i,j) =
\alpha_p S_p(i,j)
+ \alpha_m S_m(i,j)
+ \alpha_i S_i(i,j)
+ \alpha_e S_e(i,j),
  \end{equation}

where $S_p$, $S_m$, and $S_i$ denote the purpose, mechanism, and implementation similarities, and $S_e$ denotes the cross-layer edge similarity.

In our experiments, we use $K=3$, $\gamma=1.25$, and $(\alpha_p,\alpha_m,\alpha_i,\alpha_e)=(0.2,0.4,0.3,0.1)$. We use $K=3$ to retain a small set of plausible concept memberships for each item rather than assigning it to only one concept. We set $\gamma=1.25$ to give core items modestly greater influence without allowing them to dominate the remaining content. The layer weights place greater emphasis on the content layers, particularly mechanism similarity, which captures how an idea works, while assigning a smaller contribution to cross-layer structure. These parameters express the relative emphasis used in our representation rather than a universal optimum.

\subsection{Set-level diversity.}
\label{diversity_def}
To measure the overall diversity of a given idea set beyond the idea pair similarity, we use the same soft concept-membership representation to measure the diversity of an idea set. We measure how much of the concept space the set covers as a proxy for idea set diversity measurement. Within each layer, the item memberships of an idea are aggregated into its concept distribution $p_l(i)$. For each concept, we then take the strongest coverage provided by any idea in the set, capped at 1, and sum across concepts. For example, if three ideas cover one concept with strengths $0.8$, $0.7$, and $0.2$, that concept contributes only $0.8$ to the set's diversity rather than $1.7$; adding another idea that covers the same concept therefore contributes little, whereas an idea that reaches a previously uncovered concept produces a larger gain. We apply the same coverage principle to cross-layer edges. Formally,

\begin{equation}
D(S) =
\sum_l \beta_l
\sum_c
\max_{i \in S} \min\left(1, p_l(i)[c]\right)
+
\beta_e
\sum_e
\max_{i \in S} \min\left(1, m_i(e)\right),
\end{equation}

where $p_l(i)[c]$ denotes idea $i$'s aggregated membership in concept $c$ at layer $l$, and $m_i(e)$ denotes its membership on cross-layer edge $e$. We use $\beta=(1,1,1,0.5)$ to give equal importance to coverage across purpose, mechanism, and implementation, while downweighting cross-layer edges so that relational structure contributes to diversity without dominating the broader concept coverage. This coverage-based formulation gives diminishing returns to redundant ideas while rewarding ideas that expand the set into previously uncovered parts of the concept space.

\section{Evaluation}

To evaluate whether our method aligns with human experts' judgments of structural similarity among design ideas, we conducted two complementary experiments comparing our method with embedding-based methods. In experiment 1 we investigate whether the similarity metric we propose can pick out a set of items that are judged as more diverse by human experts than an embedding-based metric on a common corpus. In experiment 2 we explore a pairwise similarity task that aims to experimentally manipulate core vs. implementation  and core vs. support mechanism to isolate the influence of these constructs.


\subsection{Experiment Set up}
\subsubsection{Design Problems and Idea Corpus}
For both experiments we used a common set of four design problems, listed in Table~\ref{tab:problems}, adapted from prior ideation studies~\cite{ma2023conceptual}. We selected these problems because they are understandable without highly specialized domain knowledge while spanning different design contexts, allowing us to examine how  representations generalize across problems. For each problem, we used GPT-5.2 to generate $100$ ideas (see prompt used in Appendix~\ref{app:prompt-idea-generation}). For our method applied in both experimen, we used GPT-5.5 to construct the idea graphs and \texttt{text-embedding-3-small} to embed the concepts used in graph construction.

\subsubsection{Baselines}

We compare our approach with whole-idea embeddings approaches that are commonly used to estimate semantic
similarity in prior work~\cite{jiang2026artificial,ma2023conceptual}.
Specifically, we use two embedding models,
\texttt{text-embedding-3-small} and
\texttt{all-MiniLM-L6-v2} that are widely used in the evaluation of idea similarity and diversity assessment~\cite{jiang2026artificial,ma2023conceptual}. For each model, we evaluate a plain baseline that
embeds the original idea text and computes cosine similarity. We also evaluate
a stronger concatenation baseline~\cite{viswanathan2024large} intended to reduce dependence on surface
wording. For this baseline GPT-5.5 extracts a short description of how each idea solves the problem, and its embedding is concatenated with the original idea embedding before cosine similarity is calculated.

\subsection{Experiment 1: Set-Level Diversity}
\label{ex1}
Experiment 1 tests for how well an approach can identify ideas that contribute genuinely new content relative to an existing idea corpus. More diverse ideas selected should be perceived by experts as adding more 'novelty' to the existing idea corpus. The $100$ original ideas were split into a $40$-idea base set, representing an existing idea corpus, and a $60$-idea candidate pool to be selected from. Starting from the same base set and candidate pool, each method then selected $10$ additional ideasthat produced the largest marginal increase in the defined diversity score(section~\ref{diversity_def}). For the embedding-based baseline in experiment 1, we embed each complete idea using \texttt{text-embedding-3-small}. At each step, we compute each candidate's cosine distance to its closest idea already in the selected set and choose the candidate with the largest minimum distance, i.e., semantically dissimilar. 

\subsection{Experiment 2: Pairwise Similarity}
\label{ex2}
To further examine whether our method can capture the nuanced differences in specific idea pairs, we designed experiment 2 tests whether the graph-based similarity measure better matches
human judgments of structural similarity than embedding-based similarity.

We hypothesize that idea embeddings may blur which parts of an idea are structurally important, they may struggle when similarity in surface wording or implementation conflicts with similarity in the core mechanism. We therefore construct six triple types that allow us to test whether a similarity measure can distinguish deeper structural similarity from similarity driven by less central or more literal features.

Each triple contains one source idea and two variants generated from that source, with the variants systematically preserving or changing the core mechanism ($M$), implementation ($I$), and supporting modules ($S$). Two triple types create conflicting cues, where one variant shares the core mechanism and the other shares either the supporting modules ($M^+S^-$ vs. $M^-S^+$) or implementation ($M^+I^-$ vs. $M^-I^+$). Three isolate a difference in one idea part: supporting modules ($M^+S^+$ vs. $M^+S^-$), implementation ($M^+I^+$ vs. $M^+I^-$), or core mechanism ($M^+I^-$ vs. $M^-I^-$). The final type is an aligned similarity check ($M^+I^+$ vs. $M^-I^-$). Here, $+$ and $-$ indicate whether each part is similar to or different from the source idea, as shown in Table~\ref{tab:alignment_by_triple_type_6types}. Figure~\ref{fig:idea_sim} provides example idea pairs.

For the data used in the study, we randomly sampled 60 of the $100$ initial ideas for each problem as source ideas. We then used an advanced LLM (Opus 4.8) to generate the required variants to ensure high idea quality. The prompt used are in Appendix~\ref{app:prompt-variant-generation}. Two authors manually reviewed all 120 generated variants for each problem to verify that they matched the intended triple categories. This process produced $60$ triples per problem. We then constructed a concept graph containing both the original ideas and the generated variants and used it to calculate pairwise idea similarity using our proposed method.

\subsection{User Evaluation}
\label{user_eva}
We conducted a study including both Experiment~1 (Section~\ref{ex1}) and Experiment~2 (Section~\ref{ex2}) to assess how well our method aligns with human judgments. Each participant completed both experiments for one randomly assigned design problem from Table~\ref{tab:problems}. Each problem was evaluated by three participants, who received the same set of ideas for that problem.

\subsubsection{Participants.}
To ensure that the collected data reflected human experts' judgments of structural idea similarity, we recruited 12 participants with engineering expertise relevant to the design tasks used in the study.
 Participants had backgrounds in robotics, electrical engineering, and related engineering fields and had at least three years of engineering
experience. They were either professional engineers from industry, or current master's/PhD students in engineering who had at least 2 years' full-time industry engineering experience. Participants were recruited through advertisements on Slack and our email list. Each session took approximately 1.5--2 hours, and participants received a \$50 Amazon gift card as compensation. 

\subsubsection{Procedure}
\begin{figure}[h]
  \centering
 \includegraphics[width=\linewidth]{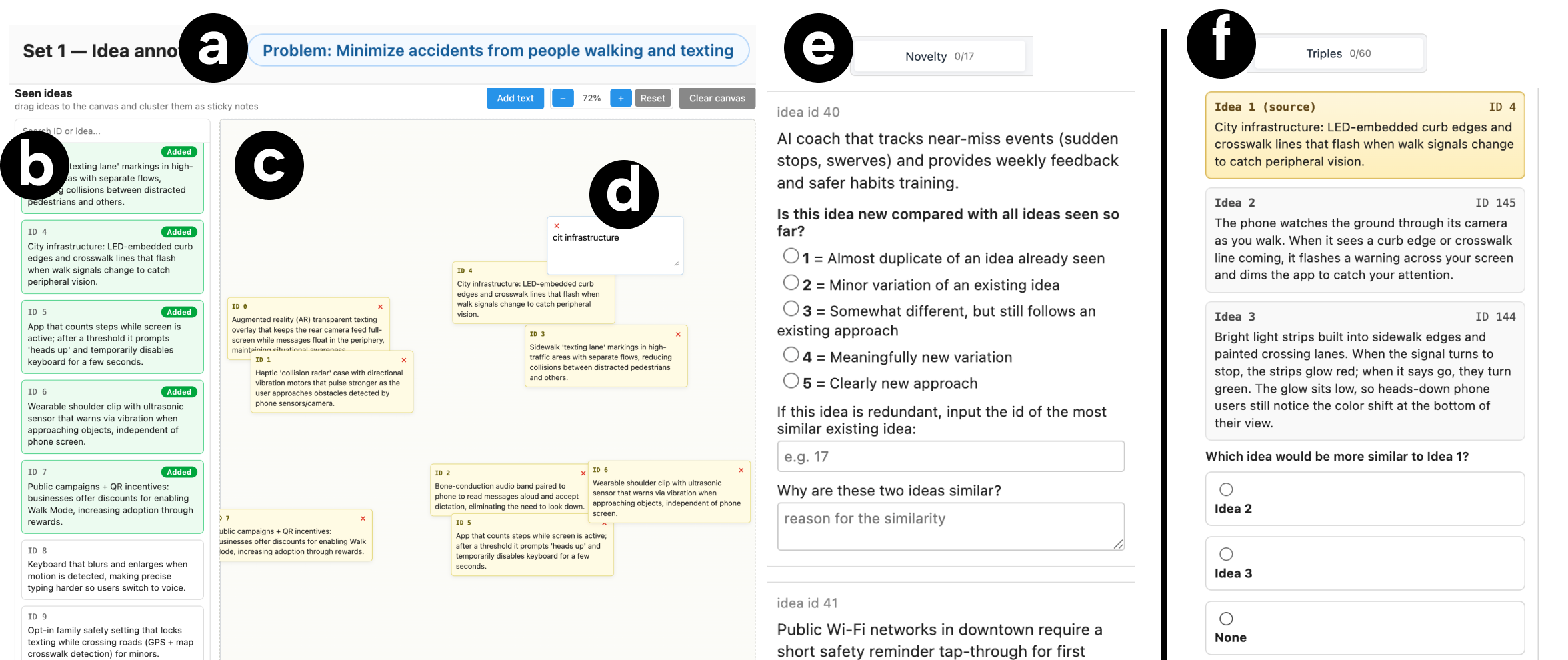}
  \caption{Annotation interface.Participants first reviewed ideas for the given design problem (a), browsed the 40 reference ideas (b), and organized them into clusters on a free-form canvas (c), with optional text notes for labeling clusters or recording observations (d). They then completed the novelty-rating task (e) and the triple-comparison task for judging which candidate idea was more similar to a source idea (f).}
  \label{fig:annotation_interface}
  \Description{}
\end{figure}

All annotation sessions were conducted online using the interface shown in Figure~\ref{fig:annotation_interface}. To help participants develop an understanding of the existing idea space before making their judgments, they were first asked to review the 40 ideas as the existing idea corpus(Figure~\ref{fig:annotation_interface}(b)) for their assigned problem (Figure~\ref{fig:annotation_interface}(a)) and organized them into groups on a canvas (Figure~\ref{fig:annotation_interface}(c)). They could optionally label these groups with notes (Figure~\ref{fig:annotation_interface}(d)).

For Experiment~1, participants rated each candidate idea on a 1--5 scale based on its novelty relative to the existing idea corpus (Figure~\ref{fig:annotation_interface}(e)), with higher scores indicating more dissimilar to the existing ideas. When they considered a candidate redundant with an existing idea, they also recorded the corresponding idea ID and briefly explained the similarity. We selected 10 ideas from the candidate pool using the embedding and our method as described in section~\ref{ex1}. Since the methods could select overlapping candidates, participants evaluated fewer than 20 unique ideas for some problems, but more than 15 in all cases.

After Experiment~1, participants completed Experiment~2 (Figure~\ref{fig:annotation_interface}(f)). Participants were asked to complete Experiment~1 first to prevent exposure to the controlled variants in Experiment~2 from influencing their novelty judgments. For Experiment 2, 
for each triple, participants viewed a source idea and two variants and selected which variant was more similar to the source, with an additional option indicating that neither was similar. Each participant evaluated 60 triples generated  as described in Section~\ref{ex2}.

\begin{table}[t]
  \caption{The four design problems used in the experiments. Each problem contributes $100$ initial ideas, one graph containing $220$ ideas, $60$ similarity triples, and one diversity-task instance.}
  \label{tab:problems}
  \begin{tabular}{l p{0.76\columnwidth}}
    \toprule
    \textbf{ID} & \textbf{Problem statement} \\
    \midrule
    1 & Design solutions for a lightweight exercise device that can be used while traveling. \\
    2 & Design solutions for a way to minimize accidents from people walking and texting on a cell phone. \\
    3 & Design solutions for a measuring cup for the blind. \\
    4 & Design solutions for a device that can help a home conserve energy. \\
    \bottomrule
  \end{tabular}
\end{table}

\section{Results}


\subsection{Experiment 1: Set Coverage Novelty}
\begin{figure}[h]
\centering
\includegraphics[width=0.8\linewidth]{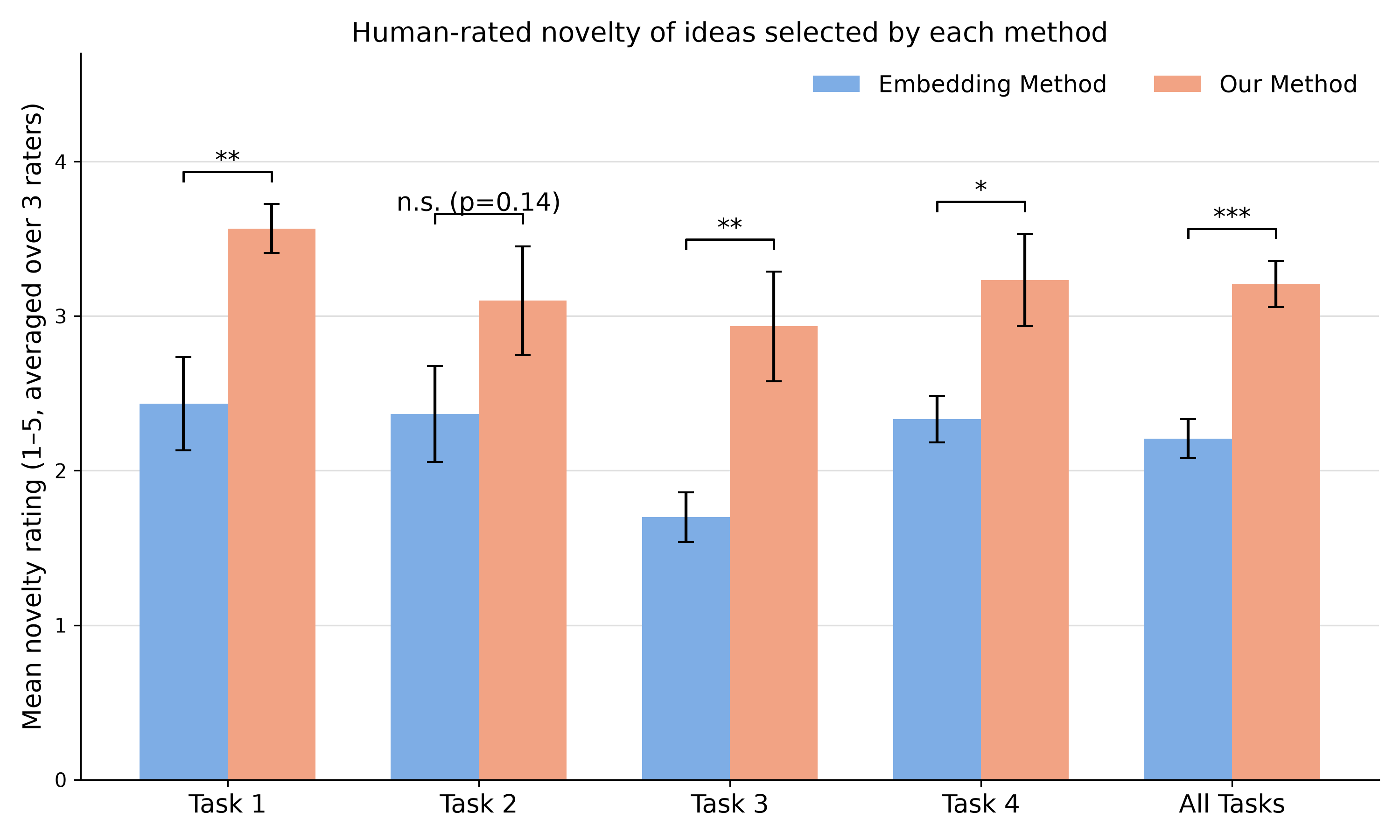}
\caption{\textbf{The graph-based method selects ideas that experts judge as more novel than those selected by the embedding-based method.} Bars show mean per-idea novelty ratings on a 1--5 scale, averaged across three independent expert raters; error bars indicate $\pm 1$ SEM. Ideas selected by both methods were excluded from this comparison. Brackets report two-sided Welch's $t$-tests (* $p<.05$, ** $p<.01$, *** $p<.001$).}
\label{fig:novelty_reu}
\Description{}
\end{figure}

Maximizing coverage with our graph-based representation selected ideas that human experts perceived as adding more novelty to the existing idea set than embedding approaches. As shown in Figure~\ref{fig:novelty_reu}, graph-selected ideas received higher human novelty ratings. Pooled across tasks, graph-selected ideas received significantly higher mean novelty ratings compared to embedding-selected ideas (3.37 vs. 2.12, two-sided Welch $t$-test on per-idea mean ratings, $t=-5.11$, $p<.001$).  Interrater reliability as measured by ICC(2,$k$) between experts across the four tasks was 0.876, indicating strong agreement in novelty judgments.

\subsection{Experiment 2: Pairwise Similarity Assessment}

\begin{table}[!t]
\centering
\small
\renewcommand{\arraystretch}{1.15}
\setlength{\tabcolsep}{6pt}
\begin{tabular}{llc}
\toprule
\textbf{Method} & \textbf{Model} & \textbf{Agreement with human annotation} \\
\midrule

Our Method
& --
& \textbf{0.871} \\

Keyphrase-concat embedding
& text-embedding-3-small
& 0.562 \\


Whole-text embedding
& text-embedding-3-small
& 0.524 \\

Keyphrase-concat embedding
& all-MiniLM-L6-v2
& 0.524 \\

Whole-text embedding
& all-MiniLM-L6-v2
& 0.511 \\

BERTScore
& --
& 0.481 \\




\bottomrule
\end{tabular}
\caption{
Accuracy of different similarity methods in aligning with the human majority
vote on the triple task, over all six triple types. Accuracy is computed over
the 233 triples whose human majority selected one of the two source-anchored
candidate ideas.
}
\label{tab:overall_alignment_6types}
\end{table}

\begin{table}[!t] \centering \small \renewcommand{\arraystretch}{1.15} \setlength{\tabcolsep}{6pt} \begin{tabular}{lcccc} \toprule \textbf{Triple type} & \textbf{$n$} & \textbf{Our Method} & \textbf{Embedding (Concat.)} & \textbf{Embedding} \\ \midrule \multicolumn{5}{l}{\textit{Conflicting cues across idea parts}} \\ \quad $M^{+}S^{-}$ vs.\ $M^{-}S^{+}$ & 33 & \textbf{0.970} & 0.788 & 0.788 \\ \quad $M^{+}I^{-}$ vs.\ $M^{-}I^{+}$ & 31 & \textbf{0.935} & 0.806 & 0.806 \\ \addlinespace[2pt] \multicolumn{5}{l}{\textit{Isolated difference in one idea part}} \\ \quad $M^{+}S^{+}$ vs.\ $M^{+}S^{-}$ & 37 & \textbf{0.811} & 0.135 & 0.054 \\ \quad $M^{+}I^{+}$ vs.\ $M^{+}I^{-}$ & 35 & \textbf{0.686} & 0.143 & 0.029 \\ \quad $M^{+}I^{-}$ vs.\ $M^{-}I^{-}$ & 36 & 0.917 & \textbf{0.972} & 0.944 \\ \addlinespace[2pt] \multicolumn{5}{l}{\textit{Aligned cues: similarity check}} \\ \quad $M^{+}I^{+}$ vs.\ $M^{-}I^{-}$ & 40 & \textbf{0.850} & 0.600 & 0.600 \\ \midrule \textbf{All} & \textbf{212} & \textbf{0.858} & 0.566 & 0.528 \\ \bottomrule \end{tabular} \caption{ Accuracy by triple type on the 212 triples for which the human majority agreed with the designed more-similar candidate. We group the triples by the type of structural comparison they require. \textit{Conflicting cues} compare candidates that each preserve a different part of the anchor idea. \textit{Isolated differences} hold one relation fixed so that the choice depends on a difference in a single idea part. The \textit{similarity check} compares a candidate that matches the anchor in both core mechanism and implementation with one that differs in both. $M$ denotes core mechanism, $S$ supporting function, and $I$ implementation; $+$ and $-$ indicate whether that part is similar to or different from the anchor. Both baselines use OpenAI text-embedding-3-small, either with a concatenated keyphrase embedding (Embedding (Concat.)) or using the idea text alone. } \label{tab:alignment_by_triple_type_6types} \end{table}

Experiment 2 aimed to directly manipulate the core structure of ideas versus the implementation or support mechanisms in order to establish greater causality about the impact of these constructs and their influence on human judgment. Overall, across triples our structure graph approach aligned substantially more with human pairwise similarity judgments than other comparison methods. Across the 233 triples for which the human majority selected one of the two source-anchored candidates, the structure graph reached 0.871 accuracy, compared with 0.562 for the strongest embedding baseline, keyphrase-concat embedding. 

To further explore the results, Table~\ref{tab:alignment_by_triple_type_6types} breaks down the triples into their generation types. To ensure the conditions represented the intended constructs, we limited this set to the 212 triples for which the human majority agreed with the synthetically generated more-similar candidate. Overall, results were similar with the full set, with our method achieving 0.858 accuracy, compared with 0.566 for Embedding (Concat.) and 0.528 for Text-emb. 

The key triples of interest were those requiring distinguishing between the core and either implementation or supporting mechanisms. In $M^{+}S^{-}$ vs.\ $M^{-}S^{+}$ triples, one candidate shared the anchor's core mechanism but differed in supporting function, while the other differed in core mechanism but shared the supporting function. Our method achieved 0.970 accuracy in these cases, compared with 0.788 for both embedding baselines. The same pattern appeared for $M^{+}I^{-}$ vs.\ $M^{-}I^{+}$ triples, where our method achieved 0.935 accuracy compared with 0.806 for both baselines. These cases provide more direct evidence than Experiment 1's set novelty task that our similarity metric distinguishes between core mechanism, supporting function, and implementation separately.

The largest advantage appeared when both candidates shared the same core mechanism and the judgment depended on a more nuanced difference in another idea part. For $M^{+}S^{+}$ vs.\ $M^{+}S^{-}$ triples, where the candidates differed in whether they preserved the anchor's supporting function, our method achieved 0.811 accuracy, compared with 0.135 for Embedding (Concat.) and 0.054 for Embedding. For $M^{+}I^{+}$ vs.\ $M^{+}I^{-}$ triples, where the candidates differed in implementation, our method achieved 0.686 accuracy, compared with 0.143 and 0.029. These results suggest that whole-idea embeddings often miss differences in a specific idea part when the candidates remain similar in their overall approach.


Our method performed most poorly when the two options differed only in implementation, achieving 0.686 accuracy on the 35 $M^{+}I^{+}$ vs.\ $M^{+}I^{-}$ triples. An error analysis revealed that the lower accuracy was largely due to all three items in a triple being clustered into the same implementation cluster, making it difficult to reliably distinguish them. This suggests that our approach could benefit from future work exploring nested clustering approaches, such as hierarchical clustering.  

Conversely, the only triple type in which the embedding baselines outperformed our method was when only one candidate preserved the core mechanism and both differed on implementation ($M^{+}I^{-}$ vs.\ $M^{-}I^{-}$). In this case our method achieved 0.917 accuracy, compared with 0.972 for Embedding (Concat.) and 0.944 for Embedding. Because a change in core mechanism creates a large difference in the overall meaning of an idea, it can often be captured directly by a whole-idea embedding.  

Finally, our method also surprisingly performed better in the similarity-check condition, where the cues across idea parts were aligned rather than conflicting. For $M^{+}I^{+}$ vs.\ $M^{-}I^{-}$ triples, one candidate shared both the core mechanism and implementation with the anchor, while the other differed in both. Our method achieved 0.850 accuracy, compared with 0.600 for both embedding baselines. The poor performance of embeddings in what should be the easiest, most different comparison was unexpected, and provides further motivation for moving to more structure-sensitive approaches.

\section{Discussion}

In this work, we explored ways to measure structual idea at scale. By separating core and supporting parts and representing purpose, mechanism, and implementation independently, \sys aligned more closely with expert judgments than the tested whole-idea embedding baselines, particularly when similarity depended on differences within a specific part of an idea. These results suggest that structural representation provides a promising basis for reasoning about similarity and diversity beyond overall semantic resemblance.

\subsection{Structural Similarity as a Foundation for Evaluating Creative AI}

Our findings suggest that evaluating idea diversity requires considering where ideas differ, rather than treating all semantic variation as equivalent.
Prior work has reached different conclusions about whether LLMs broaden or homogenize creative output, with some reporting substantial diversity and others finding convergence across generated ideas~\cite{meincke2024prompting,feng2025enhancing,anderson2024homogenization,doshi2024generative,ma2024diverse,wenger2026homogeneously}. Our results suggest that the choice of similarity measure is one factor worth considering when interpreting these findings. Whole-idea embeddings do not always align with experts' judgment of idea similarity well. A complementary experiment in the appendix~\ref{app:prompt-novelty-judge} further suggests that simply instructing an LLM judge to attend to purpose, mechanism, and implementation is not equivalent to explicitly representing those dimensions in a shared idea space: the structurally prompted judge did not show the same qualitative advantage as our representation-based approach. These findings do not explain differences across prior studies, which also vary in models, prompts, tasks, and experimental settings, but they provide a way to examine more precisely what kinds of variation are being counted as diversity.

Structural measures could therefore better support more diagnostic benchmarks for creative AI by characterizing what kinds of diversity models actually produce.
Existing work has developed automated measures and benchmarks for originality and diversity~\cite{beaty2022semantic,shypula2025evaluating,zhang2025noveltybench,guo2025benchmarking}; a structural representation could complement these approaches by separately examining variation in purposes, mechanisms, and implementations. Such benchmarks could reveal whether a new model or generation strategy expands the range of underlying approaches or mainly produces alternative realizations of familiar ones. They could also provide a more targeted way to evaluate interventions intended to increase diversity, including prompting strategies, retrieval, model combinations, or multi-stage generation pipelines.

\subsection{Structural Representations for Navigating and Exploring Idea Spaces}

Representing different forms of similarity explicitly could also support creativity tools that help people navigate idea spaces in more deliberate ways.
Creativity support systems aim to help people explore possibilities and move beyond existing directions~\cite{shneiderman2007creativity,deterding2017mixed}, yet operations such as clustering, retrieval, deduplication, or simple prompting LLM to generate`` something different'' depend on what the system considers similar. With a structural representation, ``different'' could become a more controllable interaction: a designer might ask for ideas that retain the same mechanism but use a different implementation, preserve the same purpose but use another mechanism, or modify supporting components without abandoning the core approach. These relationships correspond to distinctions that experts attended to in our evaluation, suggesting that similarity can potentially become an interaction primitive rather than only an evaluation score.

More broadly, these distinctions suggest that idea spaces could be navigated as structured families of related approaches rather than only as points in a semantic space.
Design theories have long distinguished higher-level working principles from lower-level embodiment and detail and treated differences at these levels differently when assessing idea variety~\cite{shah2003metrics,mcteague2018insights}. Extending this view, an interface might expose where ideas share a common mechanism and where they branch into different implementations, analogous to navigating a mechanism hierarchy. Structured concept spaces have already been used to support design retrieval and exploration~\cite{sarica2020technet,sweed2025finding}; explicitly representing core, supporting, and abstraction-level relationships could provide additional ways for users to move between refinement of an existing direction and exploration of structurally different ones. Whether such interactions actually improve exploration or reduce fixation remains an open empirical question.

\subsection{Open Challenges in Representing Idea Structure}

Our results also show that the value of structural similarity depends on how accurately the structure and its level of granularity are represented.
The most difficult nuanced condition for \sys involved implementation differences, and several failures occurred because concept clusters grouped implementations that experts considered meaningfully different. This points to an important tradeoff: clusters that are too fine may separate equivalent concepts expressed differently, while clusters that are too coarse may erase meaningful distinctions. Future work could explore hierarchical or progressive clustering, adaptive thresholds, or multiple levels of concept resolution so that comparisons can move between broad mechanism families and finer implementation distinctions.

The purpose--mechanism--implementation graph should therefore be viewed as one operationalization of idea structure rather than a complete representation of it.
These levels are grounded in recurring distinctions in design representations~\cite{qian1996function,shah2003metrics,mcteague2018insights}, but mechanisms themselves can exist at multiple levels of abstraction, and ideas may share a higher-level causal principle while diverging in lower-level submechanisms. More generally, structural similarity can depend on relationships among components rather than on their presence alone, as theories of structural alignment have long emphasized~\cite{gentner1983structure,gentner1993roles,markman1993splitting}. This also highlights a limitation of collapsing an entire idea into a single embedding: such compression is useful for scalable semantic matching, but can obscure compositional and relational distinctions~\cite{fodor2025compositionality,nikolaev2023representation}. Rather than replacing embeddings, future work could combine their scalability with richer hierarchical or relational representations and test how these alternatives generalize across domains and naturally occurring idea spaces.

\section{Limitations and Future Work}

Our evaluation covers four design problems and three experts per problem. Although expert judgments were consistent within the current study, a larger and more diverse set of tasks and experts is needed to determine how well the representation generalizes. In particular, the importance of purpose, mechanism, and implementation may differ across domains, especially for problems that require specialized technical knowledge. Future work should therefore evaluate the method in a wider range of design and non-design domains, recruit larger groups of domain experts, and examine whether experts with different backgrounds place different importance on the structural components of an idea.

The method also depends on how ideas are decomposed and organized into the shared concept space. The current purpose--mechanism--implementation representation is designed for solution ideas, and errors or inconsistencies in the LLM decomposition can propagate to the resulting similarity and coverage measures. Future work should test the stability of the representation across decomposition models, prompts, and graph configurations, and examine whether other domains require different structural representations or learned weights between components. Finally, the current experiments primarily establish the representation as a measurement method. The controlled similarity triples help isolate specific structural differences but do not capture all of the variation in naturally occurring idea sets. The proposed uses for diagnosing LLM homogenization and identifying unexplored combinations should therefore be tested directly on naturally generated idea spaces and in studies where people or models use these signals to generate new ideas.

\section{Conclusion}
We introduced \sys, a scalable method for measuring idea similarity and diversity through explicit structural representations of purpose, mechanism, and implementation. Across pairwise similarity and set-level coverage evaluations, \sys aligned more closely with expert judgments than whole-text embedding baselines, improving alignment with expert structural similarity judgments by 31\%. This advantage was especially pronounced when similarity depended on distinctions among different parts of an idea rather than overall semantic resemblance. By making these distinctions explicit, our work takes a step toward a stronger foundation for scalable and robust similarity metrics that can support more reliable measurement of novelty and diversity, guide idea generation toward underexplored directions, and enable creativity support tools to help people more effectively navigate and develop large idea spaces.

\section{Acknowledgments}
This research is partially with support from Toyota research institute, Google.org and the Google Cloud Research Credits program for the Gemini Academic Program.

\bibliographystyle{plainnat}
\bibliography{sample-base}

\ccsdesc[500]{Human-centered computing~Interactive systems and tools}

\appendix
\section{Appendix}
\subsection{Initial Idea Generation}
\label{app:prompt-idea-generation}
The following prompt generates the initial pool of solution ideas for each
design problem. We instantiate it with
\texttt{\{num\_solutions\}} $= 100$ and
\texttt{\{max\_id\}} $= 99$.

\begin{promptlisting}
You are a creative domain expert for the given problem. Given the following problem description, generate {num_solutions} distinct solutions.

Problem:
{problem_description}

Don't name the solution, just give a textual description of the idea, which should be a concise yet comprehensive description of the solution, including its key features and how it addresses the problem.
Return your answer strictly in the following JSON format inside <json></json> tags. Each idea should have an integer "id" starting from 0 and an "idea" field describing the solution in detail.
{output format}

Make sure you generate exactly {num_solutions} ideas, numbered from 0 to {max_id}.
\end{promptlisting}

\subsection{Variant Generation for the Triple Task}
\label{app:prompt-variant-generation}

Variants for the triple task are generated from a single shared prompt
template. Each of the four generation cases instantiates the template with
(i) two category labels \texttt{\{category\_1\_label\}} and
\texttt{\{category\_2\_label\}}, (ii) a case-specific
\texttt{\{variant\_definition\}} block that defines the required structural
relationship between the source idea and each variant category, and (iii) a
case-specific worked \texttt{\{few\_shot\_example\}} following the same
two-category pattern (omitted below for brevity). The
\texttt{\{controlled\_dimension\}} placeholder names the structural dimension
the case controls (core mechanism, implementation, or supporting mechanisms).

\subsubsection{Shared template}

\begin{promptlisting}
## Task
Generate {ideas_per_category} variants per category for the given solution and problem, for {total_idea_number} total variants.
Split the variants into two categories:
- Generate exactly {ideas_per_category} variants with "category": "{category_1_label}"
- Generate exactly {ideas_per_category} variants with "category": "{category_2_label}"

## Design Problem
{problem}

## Current Solution
{idea}

## Generation Requirements

{variant_definition}

5. Follow the requirements: 
   - Cross the wording: variants in category "{category_1_label}" (the similar category) should be written with vocabulary, phrasing, and sentence structure that differ strongly from the current solution -- use different nouns for the same components, different verbs, and different ordering of information -- even though the described {controlled_dimension} is essentially the same.
   - Conversely, variants in category "{category_2_label}" (the different category) should reuse much of the surface vocabulary, phrasing rhythm, and sentence structure of the current solution -- echo its wording where natural -- even though the described {controlled_dimension} is genuinely different.
   - Never let wording changes alter the actual mechanism or process described. The functional content, not the phrasing, must determine the category.
   - Avoid repeating the same sentence pattern across ideas within a category.

6. Write every variant in plain, readable language.
   - Keep each "idea" within 50 words.
   - Use short sentences and everyday words. Avoid jargon, buzzwords, and long technical noun phrases.

## Few-shot Example

{few_shot_example}

## Output Format

Return your answer strictly in the following JSON format inside <json></json> tags.

Each idea should include:
- "id": an integer starting from 0
- "category": one of "{category_1_label}" or "{category_2_label}"
- "idea": a plain-language description of the solution within 50 words, stating what it is and how it solves the problem
- "explanation": a short explanation of why this idea is a good variant according to the requirements
Do not include explanations outside the JSON.
The example below shows the JSON structure only; return all {total_idea_number} requested ideas.

{output format}
\end{promptlisting}

\subsubsection{Case 1: same core mechanism, similar vs.\ different implementation}

Category labels: \texttt{same\_core\_same\_implementation},
\texttt{same\_core\_different\_implementation}.
The \texttt{\{variant\_definition\}} block:

\begin{promptlisting}
1. The generated variants should be functionally close to the current solution.
   - All variants, together with the original solution, should share a similar underlying core mechanism.
   - The core mechanism means the main functional principle that makes the solution work, not the specific device, material, interface, or technology used.

2. For category "{category_1_label}", keep the implementation of the core mechanism similar to the current solution.
   - The variant should preserve essentially the same implementation style, device setup, material choice, interface mode, or system architecture as the current solution.
   - It may still vary details such as geometry, control logic, operating sequence, scale, or user workflow.
   - Do not simply copy the current solution verbatim. It should still read like a distinct concept.

3. For category "{category_2_label}", change the implementation of the core mechanism.
   - Each variant should realize the same or highly similar core mechanism through a different technical, material, interaction, or system implementation.
   - Do not simply rename the same solution or make only superficial changes.

4. The supporting modules should remain similar across variants.
   - Supporting modules are secondary functions that help the main mechanism work, such as containment, activation, state transition, separation, filtering, recovery, reuse, feedback, or mode switching.
   - The generated variants should use comparable supporting functions, but they may implement or describe them differently.
\end{promptlisting}

\subsubsection{Case 2: different core mechanism, similar vs.\ different implementation}

Category labels: \texttt{different\_core\_same\_implementation},
\texttt{different\_core\_different\_implementation}.
The \texttt{\{variant\_definition\}} block:

\begin{promptlisting}
1. For category "{category_1_label}", the generated variants should share a similar implementation style with the current solution.
   - The implementation means the concrete device setup, material format, interaction mode, system architecture, or operating procedure used to realize the solution.
   - The variants may use similar components, environments, or process structures as the current solution.

2. The underlying core mechanisms should be different across variants.
   - The core mechanism means the main functional principle that makes the solution work.
   - Each variant should solve the problem through a different functional principle, even if it uses a similar device, medium, interface, or setup.
   - Do not generate variants that only make small technical changes while preserving the same main mechanism.

3. For category "{category_2_label}", the generated variants should also use a different implementation style from the current solution.
   - Change the concrete device setup, material format, interaction mode, system architecture, or operating procedure, not just the main functional principle.
   - The variant should still address the same problem effectively, but it should no longer look like the same implementation family.

4. The supporting modules should remain similar across variants.
   - Supporting modules are secondary functions that help the solution work, such as containment, activation, delivery, separation, removal, recovery, reuse, feedback, or mode switching.
   - The generated variants should use comparable supporting functions, but they may implement or describe them differently.
\end{promptlisting}

\subsubsection{Case 3: same core mechanism, similar vs.\ different supporting mechanisms}

Category labels: \texttt{same\_core\_same\_supporting\_mechanisms},
\texttt{same\_core\_different\_supporting\_mechanisms}.
The \texttt{\{variant\_definition\}} block:

\begin{promptlisting}
1. The generated variants should share a similar underlying core mechanism with the current solution.
   - The core mechanism means the main functional principle that makes the solution work.
   - All variants, together with the original solution, should solve the problem through the same or highly similar main mechanism.
   - Do not generate variants that only look similar on the surface but rely on a different functional principle.

2. For category "{category_1_label}", keep the supporting mechanisms similar to the current solution.
   - Supporting mechanisms are secondary processes that help deliver, activate, control, remove, recover, or complete the main mechanism.
   - The generated variants should preserve comparable delivery steps, activation conditions, operating flow, residue handling, recovery steps, or user workflow.
   - The variant may still change implementation details, but the support structure should feel recognizably similar.

3. For category "{category_2_label}", the supporting mechanisms should be different from the current solution.
   - Supporting mechanisms are secondary processes that help deliver, activate, control, remove, recover, or complete the main mechanism.
   - Each variant should support the shared core mechanism through a different system setup, delivery method, activation condition, removal process, operating environment, or user interaction.
   - The difference should be more than wording. The variant should change how the supporting functions are carried out.

4. The implementation details may also differ, but the main emphasis should be on controlling whether the supporting mechanisms stay similar or become different.
\end{promptlisting}

\subsubsection{Case 4: different core mechanism, similar vs.\ different supporting mechanisms}

Category labels: \texttt{different\_core\_same\_supporting\_mechanisms},
\texttt{different\_core\_different\_supporting\_mechanisms}.
The \texttt{\{variant\_definition\}} block:

\begin{promptlisting}
1. The generated variants should have different underlying core mechanisms from the current solution.
   - The core mechanism means the main functional principle that makes the solution work.
   - Each variant should solve the problem through a different main mechanism, not just a different material, ingredient, or device name.
   - Do not generate variants that only rephrase the same mechanism or make superficial substitutions.

2. For category "{category_1_label}", the supporting mechanisms should be similar to the current solution.
   - Supporting mechanisms are secondary processes that help deliver, activate, control, remove, recover, or complete the main mechanism.
   - The generated variants should keep a similar support structure, such as similar delivery method, operating environment, activation condition, removal process, recovery step, or user workflow.
   - The supporting mechanisms do not need to be identical, but they should play comparable functional roles.

3. For category "{category_2_label}", the supporting mechanisms should be different from the current solution.
   - Change the delivery method, operating environment, activation condition, removal process, recovery step, or user workflow instead of keeping the same support scaffold.
   - The variant should still solve the problem through a different core mechanism, and its support process should also feel materially different.

4. The implementation style may be similar or different, but the key distinction should come from the changed core mechanism plus the requested supporting-mechanism relationship.
   - For example, two ideas may both use a mist-based chamber, humidity or temperature control, and residue removal.
   - However, one may remove dirt through micro-mechanical deformation, while the other may remove dirt through chemical decomposition.
\end{promptlisting}

\subsection{Idea Decomposition}
\label{app:prompt-decomposition}

The following prompt decomposes each idea into the three structural layers
(purposes, mechanisms, implementations) used to build the concept graph.

\begin{promptlisting}
## Goal
You are an expert in the domain of the given problem. You are trying to determine the similarity of solutions to the given problem based on how they solve it. To do this, you are now decomposing each idea to reveal its essence for later comparison.

## Task
Decompose the given idea into three decoupled layers:
- **Purpose** -- sub-problems the idea solves, stated as *goals* (not methods). Phrased generically enough to apply to alternative solutions for the same problem. Focus on the purpose modules directly related to solving the problem and ignore those not relevant but mentioned in the idea.
- **Mechanisms** -- the causal effects that make each module achievable. Each mechanism is given in two forms: a `principle` (concrete) and an `abstract` (role-generic).
- **Implementations** -- the concrete devices, materials, techniques, or procedures used.

## Rules
### Shared requirements
- Each item is minimal -- no "and"/"or" joining distinct concepts.
- No overlap within a layer or across layers.
- Cover every distinct purpose/mechanism/implementation mentioned in the idea.
- The three layers are decoupled: one mechanism may serve multiple purposes, one implementation may realize multiple mechanisms, one purpose may rely on multiple mechanisms.
- Mark the single core purpose and single core mechanism doing the central work.

### Layer-specific rules
**Purpose**
- State as goals, not methods. Do not restate the original problem or use similar expressions as a module.
- Use vocabulary that could equally describe other solutions to the same problem.
- Focus on the main procedures and subgoals for solving the problem, and ignore those irrelevant but mentioned in the idea.
- **For each component, check if removed, would the remaining components lose a meaningful functional step, or only just lose detail that won't affect the overall idea structure?**

**Mechanisms**
- Each mechanism describes a *single* cause-effect step. Split any "and then" or "which causes" into separate mechanisms.
- Provide two forms:
  - `principle`: the effect in concrete language drawn from the idea.
  - `abstract`: the same effect with every concrete noun replaced by a role word (body, medium, interface, energy source, agent, signal, etc.). **Test**: for each noun in `principle`, ask *"would this sentence still be true if a different physical thing played this role?"* If yes, keep it; if no, abstract it.
- A mechanism is never a goal and never a device. Prefer short noun phrases to express the effect.
- **Avoid overlaps between Mechanisms and Purposes**

**Implementations**
- One concrete device/material/technique/procedure per item. Never merge multiple things using "and"/"or".

**Relations between purposes**
- Exactly two purposes per relation, at most one relation per pair.
- Active-voice verb phrase. Only include strong relations that directly support the idea's effectiveness.
- Directional (causal, conditional, sequential): source first in `module_ids`. Symmetric (synergy, complementary): mark `undirectional`.

## Example
{Example}

## Input
problem: <<{problem}>>
idea: <<{idea}>>

## Instructions
1. Filter the redundant info in the idea to shorten the expression to get the essence of the idea.
2. Identify the components according to the rules above and the processed idea in step 1.
3. Stress test by paraphrasing the idea and checking if the same components are still valid. If not, adjust the components accordingly to make them more robust.
4. Check if the components meet the requirements. If not, adjust the results accordingly.

## Output: give answers of following questions according to the steps above.
Only output answers without repeating questions using the format:
1. answers for Q1
...
2. answers for Q4

Q1: what is the short version of the idea in step 1?
Q2: what are the components extracted in step 2?
Q3: what is your paraphrased version of the idea in step 3? If doing extraction on this version, would you get the same components? If not, adjust the components to make them more robust and explain how you adjust them.
Q4: Output the final results using the following format: Ids start at 0 within each list. Enclose the final answer in `<json>` tags. Do not add code fences.
{output format}
\end{promptlisting}

\subsection{LLM Judge for Novelty Selection}
\label{app:prompt-novelty-judge}

\subsubsection{Comparison with LLM-Based Selection}

Directly prompting an LLM to assess idea novelty did not identify sets with higher human-rated novelty than the graph-based method. We tested two prompting conditions: a \emph{plain LLM prompt}, which directly asked the LLM to assess novelty, and a \emph{structure-informed LLM prompt}, which additionally provided our definition of idea structure and the rules used to compare similarity between ideas (see the two following sections).

First, we applied both prompts to the ideas selected by the graph-based and embedding-based methods. For each prompt, we formed a top-10 set per problem in two ways: using the LLM's 1--5 novelty ratings and using its direct novelty ranking. We then evaluated these selected sets using the same human novelty ratings used in the main experiment, allowing all selection methods to be compared using a common measure.

As shown in Table~\ref{tab:novelty_judge_top10_incl_both}, all LLM-based selections had lower mean human novelty ratings than the graph-based selection. The graph-based method achieved a mean human novelty rating of 3.208, whereas the LLM-based selections ranged from 2.858 to 2.942. These differences were not statistically significant in pairwise Welch's $t$-tests ($p=.136$--$.243$). The LLM-selected sets also overlapped with both graph- and embedding-selected ideas, rather than consistently recovering the graph-selected ideas. These results suggest that directly prompting an LLM to assess novelty does not reproduce the selection produced by the graph-based coverage measure.

Second, directly asking the LLM to select ideas from the full candidate pool produced substantially different selections from both diversity-based methods. Given the same 40 base ideas and 60 candidate ideas, each prompt selected 10 additional ideas per problem. Across the four problems, the plain LLM prompt selected 10 of the 40 graph-selected ideas (25.0\%) and 6 of the 40 embedding-selected ideas (15.0\%). The structure-informed prompt selected 10 of the 40 graph-selected ideas (25.0\%) and 9 of the 40 embedding-selected ideas (22.5\%). Thus, adding the structural definition and similarity comparison rules did not substantially increase agreement with the graph-based method. The low overlap suggests that the graph-based coverage measure selects ideas differently from direct LLM prompting, even when the LLM is given the same structural principles used by our method.

\begin{table}[!t]
\centering
\small
\renewcommand{\arraystretch}{1.15}
\setlength{\tabcolsep}{6pt}
\resizebox{\linewidth}{!}{%
\begin{tabular}{llccc}
\toprule
\textbf{Method}
& \textbf{Selection criterion}
& \textbf{Coverage of human rated top-50\%}
& \textbf{Mean human novelty}
& \textbf{Welch $p$ vs.\ graph} \\
\midrule

Our Method
& coverage gain (our method)
& \textbf{32/40 (80.0\%)}
& \textbf{3.208}
& -- \\

Embedding
& max-min cosine distance
& 9/40 (22.5\%)
& 2.208
& $<$0.001 \\

LLM (plain prompt)
& own 1--5 novelty rating
& 23/40 (57.5\%)
& 2.858
& 0.136 \\

LLM (w/ structural instruction)
& own 1--5 novelty rating
& 25/40 (62.5\%)
& 2.942
& 0.243 \\

\bottomrule
\end{tabular}%
}
\caption{
Mean human novelty rating of the top-10 ideas selected per problem by each
method, over four problems (40 selected ideas per method). The two LLM judges (plain and
structural prompts, gpt-5.5) select their top 10 from the same candidate pool
either by their own 1--5 novelty rating. Coverage of human top-50\% counts,
over each method's 40 selected ideas, how many fall in the top half of their
problem's candidate pool by mean human novelty rating.
The last column is the two-sided Welch $t$-test $p$-value comparing the
method's selected-idea human novelty scores against the graph selection's.}
\label{tab:novelty_judge_top10_incl_both}
\end{table}

\subsubsection{Structural style instruction}

\begin{promptlisting}
When you compare ideas, judge them by their STRUCTURAL representation: the purpose (what problem aspect it addresses), the core working principle (the main functional mechanism that makes it work), and the concrete implementation (device, material, process).
Steps for analyzing each idea:
1. Identify the mechanism blocks in the idea; each block has one high-level purpose, one mechanism, and its specific implementation.
2. Identify the core mechanism block among the blocks, and put more weight on this core when comparing similarity.
3. When comparing, focus more on the working principle than on the implementation details.
\end{promptlisting}

\subsubsection{Novelty rating prompt}

\begin{promptlisting}
You are judging the novelty of a new solution idea for the design problem: "{problem}".

{style_instruction}## Existing ideas (the reference set)
{base_ideas}

## New idea to score
{candidate_idea}

## Task
Rate how novel the new idea is relative to the existing reference set, on a 1-5 scale:
1 -- Almost duplicate of an idea already seen
2 -- Minor variation of an existing idea
3 -- Somewhat different, but still follows an existing approach
4 -- Meaningfully new variation
5 -- Clearly new approach

Also report the id of the existing idea that is most similar to the new idea (null if nothing is meaningfully similar).

Answer strictly in JSON inside <json></json> tags:
{output format}
\end{promptlisting}
\end{document}